\documentclass[prl,twocolumn,showpacs,preprintnumbers,amsmath,amssymb,nofootinbib]{revtex4}

\usepackage{bbm,bm,slashed}
\usepackage{graphicx}
\usepackage{color,array,tabularx}

\definecolor{carsten}{rgb}{0.9,0,0}

\definecolor{stefan}{rgb}{0.9,0,0.9}

\definecolor{michael}{rgb}{0,0.6,0}

\begin{document}

\title{Instabilities of interacting electrons on the honeycomb bilayer
} \date{December 22, 2011} \author{Michael M. Scherer}
\email{scherer@physik.rwth-aachen.de}
\author{Stefan Uebelacker}
\author{Carsten Honerkamp}

\pacs{71.10.-w, 73.22.Pr}

\affiliation{Institute for Theoretical Solid State Physics, RWTH Aachen University, D-52056 Aachen  \\  and JARA - FIT Fundamentals of Future Information Technology 
}

\begin{abstract} 
We investigate the instabilities of interacting electrons on the honeycomb bilayer by means of the functional renormalization group for a range of interactions up to the third-nearest neighbor. 
Besides a novel instability toward a gapless charge-density wave we find that
using interaction parameters as determined by ab-initio calculations for graphene and graphite puts the system close to the boundary between antiferromagnetic and quantum spin Hall instabilities. Importantly, the energy scales for these instabilities are large such that imperfections and deviations from the basic model are expected to play a major role in real bilayer graphene, where interaction effects seem to be seen only at smaller scales. We therefore analyze how reducing the critical scale and small doping of the layers affect the instabilities.
\end{abstract}

\maketitle
\begin{figure}
 \includegraphics[width=0.48\textwidth]{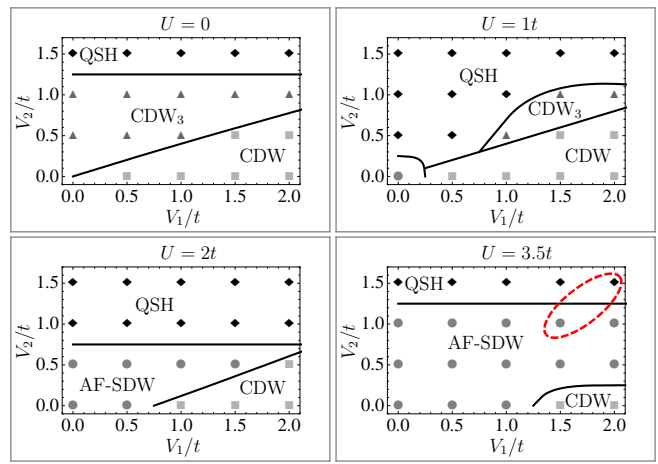}
\caption{Tentative fRG phase diagrams for $U/t=0,1,2,3.5$ and $t_\perp=0.1t$. The area encircled by the dashed line shows the region of the cRPA parameters of Ref. \cite{wehling2011}.}
\label{fig:Pd01}
\end{figure}
\begin{figure}[t!]
\centering
\includegraphics[width=0.19\textwidth]{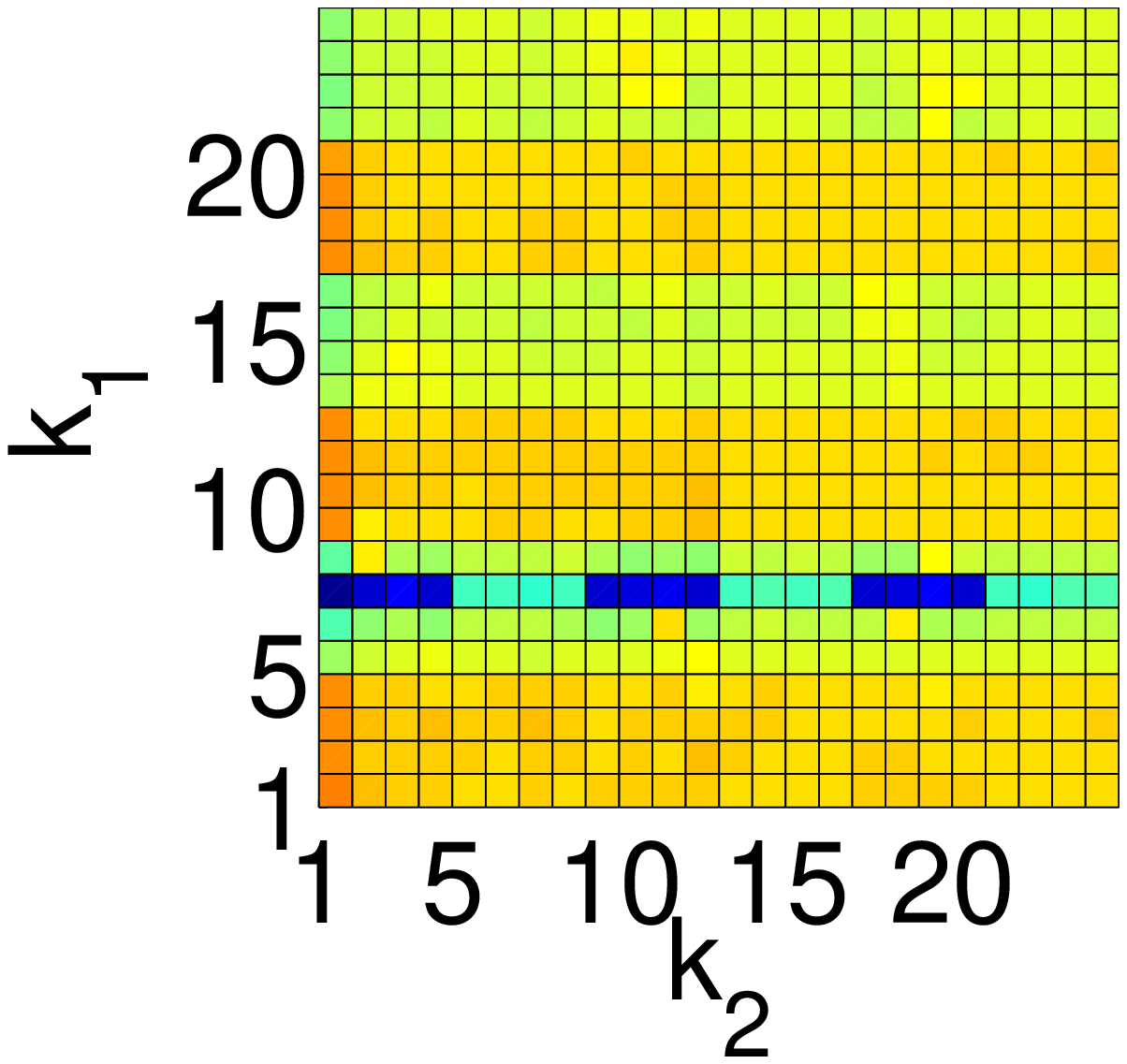}\hspace{-0.7cm}
\includegraphics[width=0.19\textwidth]{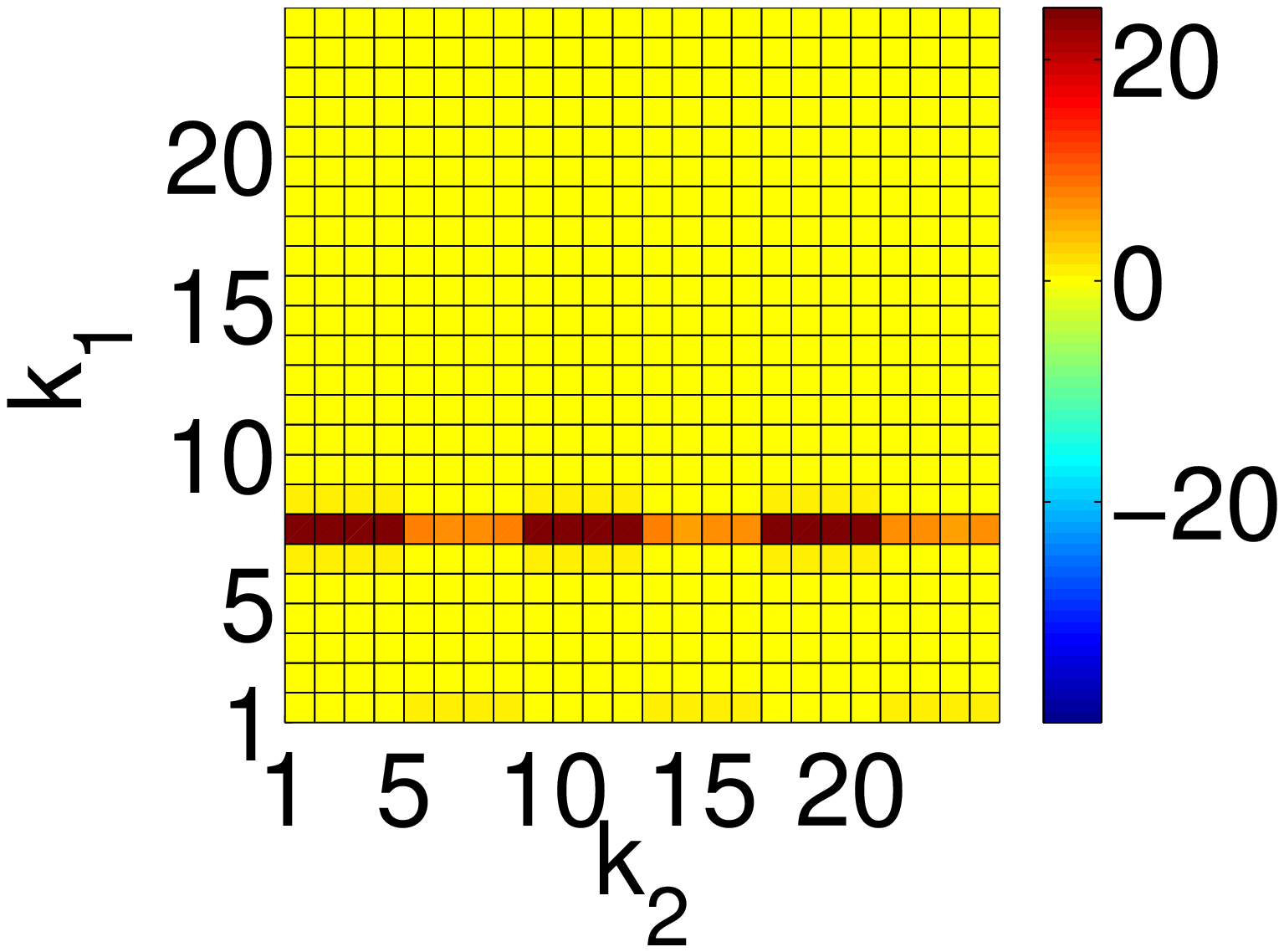}
\includegraphics[width=0.13\textwidth]{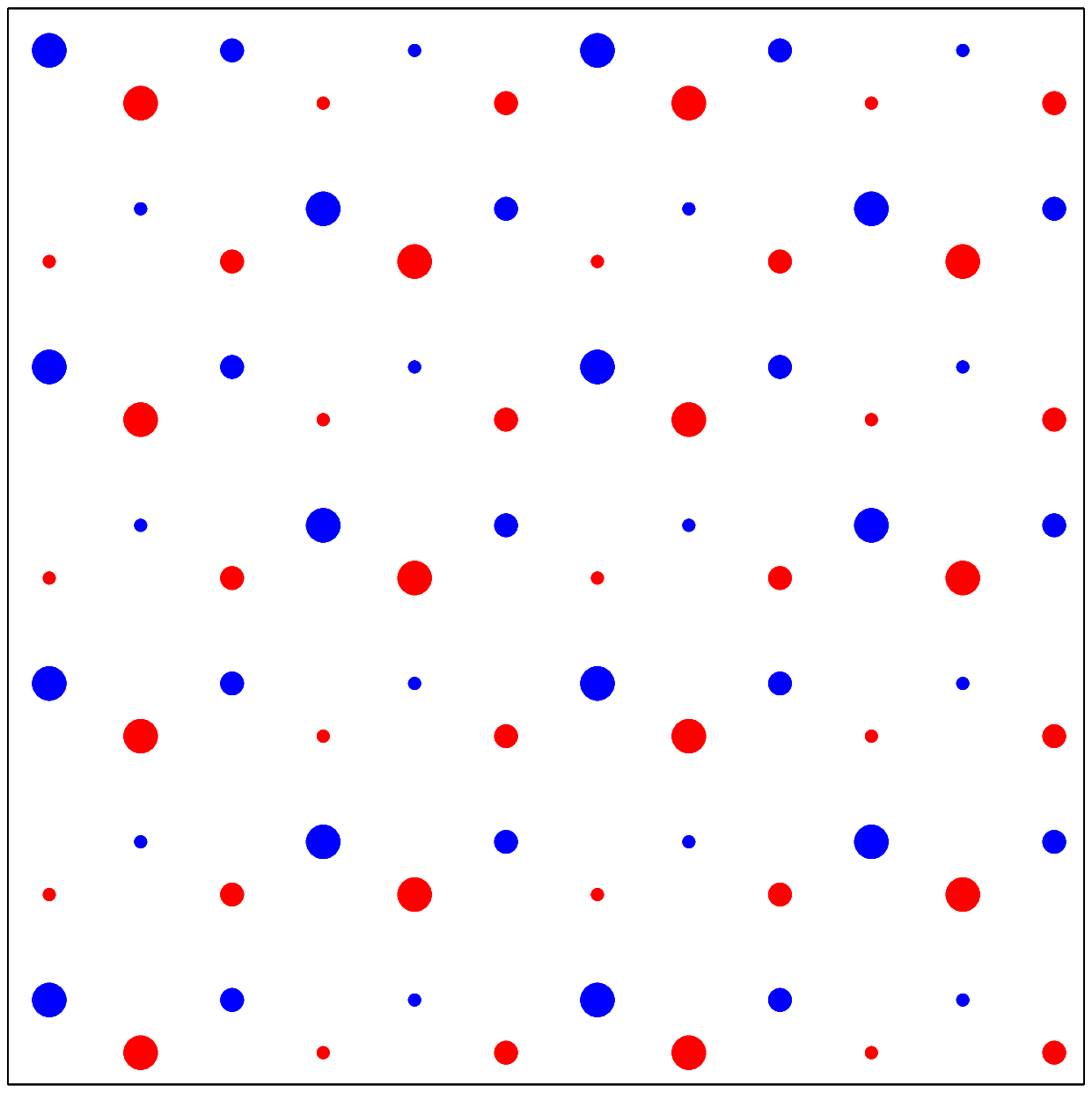}
\caption{CDW$_3$ instability. Left \& middle plot: Effective coupling in units of $t$ versus incoming patch indices $k_1$ and $k_2$ for $U=0,V_1=0.5t,V_2=0.5t$ and $t_\perp=0.1t$, with first outgoing wavevector $k_3$ on patch 1. Left plot:  $o_1=o_2=o_3=o_4$.  Middle plot: $o_1=o_3$ and $o_2=o_4\neq o_1$. Right: Qualitative charge distribution in the CDW$_3$ mean-field state for $\alpha=1/6$ and equal $|\Delta_o|$s in one layer. The size of the symbols indicates the charge density, the three sizes in one (original) sublattice average to 1.}
\label{fig:realspace-mcdw}
\end{figure}

Recently, a number of experiments have addressed the ground state properties of bilayer graphene\cite{exp01,exp02,exp03,exp04}, trying to clarify possible signatures for the interaction between the electrons. 
So far, however, the precise nature of the ground state and whether the spectrum is gapped or gapless remain under controversial discussions.  A number of competing instabilities of the semi-metal, e.g. toward a gapped antiferromagnetic (AF) state, a gapless nematic phase and even toward gapped topological phases as the quantum anomalous and quantum spin hall states are debated and compared to the experimental findings. These proposals are subject of numerous theoretical works\cite{nilsson2006,mccann2007,hongki2008,lemonik2010,vafek2010a,nandkishore2010,zhang2010,vafek2010b,vafek2011,khari2011}. Comprehensive renormalization group (RG) investigations of the different interaction processes giving rise to the various instabilities have been performed\cite{vafek2010b,vafek2011} within a continuum model for the vicinity of the band crossing points (BCPs), also addressing the role of the range of the interactions. These studies indicate that the ground state properties may depend decisively on the profile of the interaction.  In this context it is important to notice that the effective interaction parameters and their spatial dependence for the usual low-energy models of graphene and graphite have been calculated by ab-initio techniques\cite{wehling2011}, using the constrained random phase approximation (cRPA) that takes into account the screening due to bands further away from the Fermi level.  
In this work we use a functional RG (fRG) scheme for an unbiased investigation of the instabilities of the A-B or (Bernal-)stacked bilayer honeycomb lattice. Besides exploring the general picture, we use the interaction parameters as specified by the ab-initio calculations\cite{wehling2011}. This allows us to get closer toward a realistic picture of the dominant ordering tendencies in bilayer graphene.
\\ \indent
We start with the model for the graphene bilayer at the charge neutrality point. With respect to a site in sublattice A in honeycomb layer $l$, the three nearest neighbor (n.n.) sites on sublattice B in the same layer  are given by  the shifts $\vec{\delta}_i$ with $i \in \{1,2,3\}$. The n.n. hopping part of the Hamiltonian within layer $l$ has the form\cite{castroneto2009}
 $H_l^{\parallel}=-t\sum_{s,\vec{R},\vec{\delta}_i}\big(b_{l,s}^\dagger(\vec{R}+(-1)^{l-1}\vec{\delta}_i)a_{l,s}(\vec{R})+\mathrm{h.c.}\big), $
where $s = \uparrow,\downarrow$ is the electron spin, $a_{l,s}(\vec{R})$ the annihilation operator of an electron at site $\vec{R}$ on sublattice A,  $b_{l,s}^\dagger$ the creation operator on sublattice B, and $l$ the layer index.  The relative sign for the neighbors $\vec{\delta}_i$ in the two layers is due to the A-B-stacking. The interlayer hopping $t_\perp$ between sites that lie on top of each other (A$_1$ and A$_2$-sites) reads\cite{castroneto2009}
$
 H^{\perp}=t_{\perp}\sum_{s,\vec{R}}\big(a_{1,s}^\dagger(\vec{R})a_{2,s}(\vec{R})+\mathrm{h.c.}\big).
$
Diagonalizing $H_\mathrm{free}= \sum_{l=1,2} H_l^{\parallel}+ H^{\perp}$ results in 4 bands. 
Two of them have two inequivalent BCPs $K$ and $K^\prime$ at the Brillouin zone (BZ) corners at the Fermi level. While $t_\perp=0$ yields two copies of the single layer with linear dispersion (Dirac cones) at $K$ and $K^\prime$, $t_\perp \neq 0$ results in a quadratic dispersion near those Fermi points. This allows interaction-driven instabilities for arbitrarily small couplings. The instabilities of a single quadratic band crossing point were analyzed in Refs.\cite{sun,uebelacker}. We ignore trigonal warping terms for the time being, but comment later on the effect of perturbations on the quadratic BCPs.
\\ \indent
As interactions, we account for an onsite repulsion $U$, a n.n. density-density interaction $V_1$ and a second-n.n. density-density interaction $V_2$. For these terms, cRPA values are listed in Ref.\cite{wehling2011}. 
In addition, 
for checking the robustness of our results, we consider a third-n.n. repulsion $V_3$.  The interaction Hamiltonian reads
\begin{eqnarray}\label{eq:longH}
 H_\mathrm{int}&=&U\sum_i n_{i,\uparrow}n_{i,\downarrow}+V_1\sum
 _{\langle i,j \rangle, s, s^\prime} 
 n_{i,s}n_{j,s^\prime}\\
 &+&V_2\sum
 _{\langle \langle i,j \rangle \rangle, s, s^\prime} 
 n_{i,s}n_{j,s^\prime}+V_3\sum
 _{\langle\langle\langle i,j \rangle\rangle\rangle, s, s^\prime} 
 n_{i,s}n_{j,s^\prime}\,.\nonumber
\end{eqnarray}
Here $i,j$ run over the lattice sites, but pairs are included only once.
The unitary transformation from the orbital fields to the bands diagonalizing $H_\mathrm{free}$ has to be carried out on $H_\mathrm{int}$ as well. This adds {\em orbital makeup} to the interaction terms in band representation, leading to an additional angular dependence of the interactions near the $K$, $K'$-points. In order to resolve this dependence we use an angular patching of the interaction terms as explained below. This way our study goes beyond the 'g-ology' approach in Ref. \cite{vafek2011}.
\\ \indent
\begin{figure}
 \includegraphics[height=0.2\textwidth]{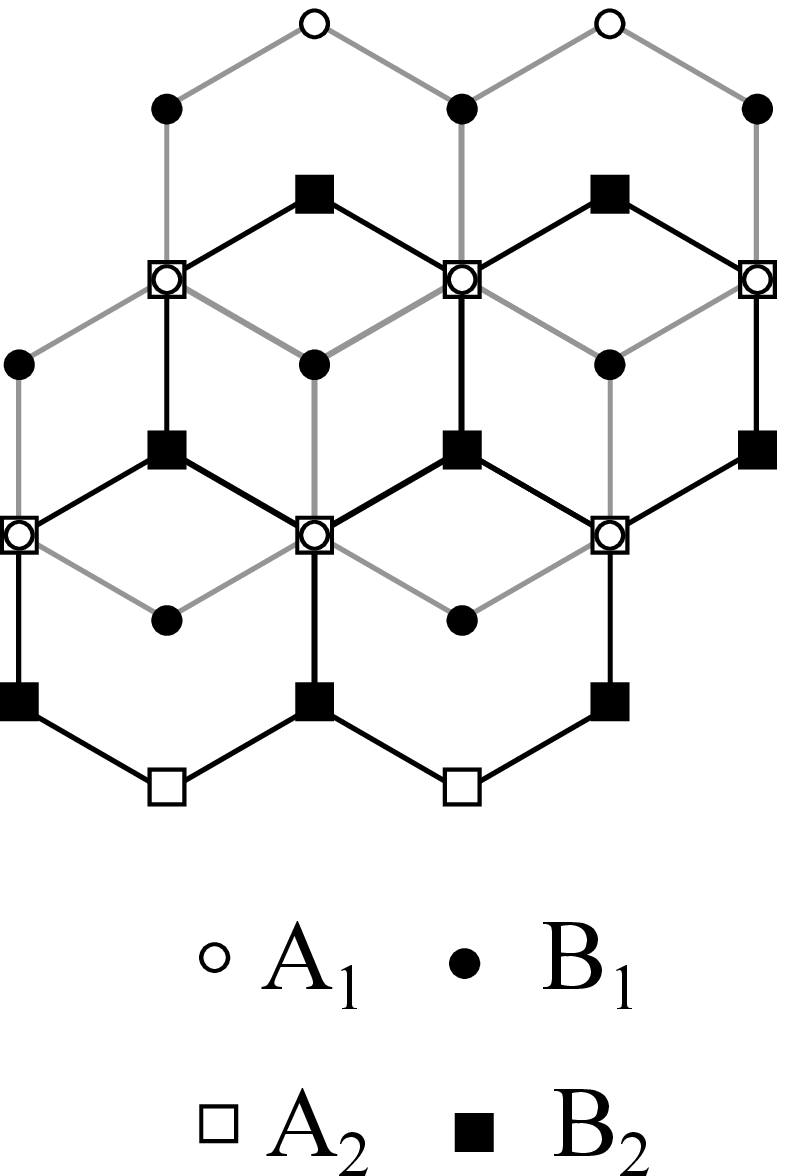}
 \hspace{0.2cm}
  \includegraphics[height=0.2\textwidth]{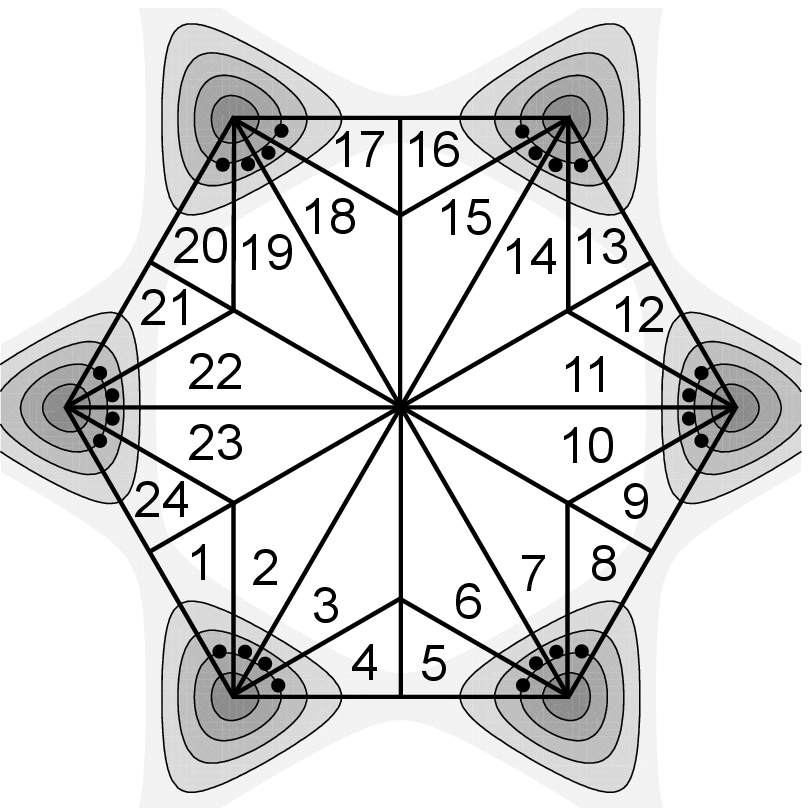}
\caption{Left panel: Sketch of Bernal-stacked bilayer graphene with the sublattices $\mathrm{A}_l, \mathrm{B}_l$ and layer index $l \in \{1,2\}$. Right panel: Patching scheme of the Brillouin zone in the fRG. The black dots denote the momentum vectors at which the coupling function is evaluated.}
\label{fig:patching} 
\end{figure}  
We employ a functional renormalization group (fRG) approach for the one-particle-irreducible vertices of a fermionic many-body system (for a recent review, see \cite{metzner2011}). In this scheme, an infrared regulator with energy  scale $\Lambda$  is introduced into the bare propagator. The RG flow is generated upon  variation of $\Lambda$. By integrating the flow down from an initial scale $\Lambda_0$  of the order of the bandwidth to the infrared $\Lambda \rightarrow 0$, one smoothly interpolates between the bare action of the system and the effective action at low energy. Here, the hierarchy of flowing vertex functions is truncated after the four-point (two-particle interaction) vertex. This vertex is described by a coupling function $V_\Lambda(k_1,k_2;k_3,k_4)$ with a discretized momentum dependence for incoming quantum numbers $k_1, k_2$ and outgoing  $k_3, k_4$. Here $k_i$ includes a wavevector $\vec{k}_i$, a Matsubara frequency $\omega_i$, a spin projection $s_i$, and for our case also band-, or orbital indices. For simplicity,  and as one is interested in groundstate properties, external frequencies are set to zero. Furthermore,  in order to keep the calculations doable, we neglect self-energy corrections. This approximate fRG scheme then amounts to an infinite-order summation of one-loop particle-particle and particle-hole terms of second order in the effective interactions. It allows for an unbiased investigation of the competition between various correlations, by analyzing the components of $V_\Lambda(k_1,k_2;k_3,k_4)$ that create instabilities by growing large at a critical scale $\Lambda_c$\cite{metzner2011}. With the approximations mentioned above, this procedure is well-controlled for small interactions. At intermediate interaction strengths we still expect to obtain reliable results. In any case, the fRG takes into account effects beyond mean-field and RPA.
The scale $\Lambda_c$ can be interpreted as estimate for ordering temperatures, if ordering is allowed, or at least as temperature below which the dominant correlations should be clearly observable. Furthermore, one can understand $\Lambda_c$ as energy scale for the modification of the spectrum, typically by a gap. \\
The discretization of the interaction vertex $V_\Lambda$  is implemented by dividing the BZ into $N$ patches with constant  wavevector-dependence within one patch. The representative momenta for the patches are chosen to lie at or close to the Fermi level. The patching scheme is shown in Fig. \ref{fig:patching}, with $N=24$. Each of the four momenta in $V_\Lambda(k_1,k_2;k_3,k_4)$ is additionally equipped with a sublattice index and a layer index. Momentum conservation fixes one of the four wavevectors. Altogether this results in a $4^4\times N^3$ component coupling function $V_\Lambda$. \\
In this work we study the flow at temperature $T=0$. We find flows to strong coupling with non-zero critical scales $\Lambda_c$ for all choices of non-vanishing interaction terms provided $t_\perp \not=0$, due to the non-vanishing density of states at the Fermi level of the coupled layers \cite{unserletter}. In practice, the flow is stopped at some finite value for the largest interaction component of the order of twice the bandwidth. Then $\Lambda_c$ is determined by extrapolation.
\\ 
Here we present fRG results at zero temperature for interlayer hopping $t_\perp =0.1t$ as estimate for values in the literature, e.g.  for graphite\cite{Zhang2008} or for few-layer graphene\cite{ohta2007}. We also analyzed larger $t_\perp<t$, without major qualitative differences. The monolayer-case $t_\perp=0$ was studied  by fRG in Refs. \cite{Honerkamp2008,HonerkampRaghu2008} and in Refs. \cite{kiesel,dhlee} for larger doping.  The next-neighbor hopping $t \approx 2.8eV$ sets the energy unit. We then study the parameter space spanned by $U, V_1$ and $V_2$ up to the cRPA parameters found in\cite{wehling2011}. By identifying the leading tendencies, i.e. the strongest class of divergent couplings, we encounter rich tentative phase diagrams shown in Fig.~\ref{fig:Pd01}. The drawn boundaries are guides to the eye. Typically, the flows change continuously from one regime to the other without drastic features in $\Lambda_c$. Hence, while without including self-energy effects a possible suppression of the critical scales due to quasiparticle degradation is not captured, we expect that these transitions are of first order. We now discuss the various ordering tendencies found for given parameters and how they are revealed in the fRG flow.
\\
{\em Antiferromagnetic (AF) spin density wave (SDW) instability. }
In the fRG data, the flow towards the AF-SDW is seen as a leading divergence of interaction components with zero momentum transfer in the spin channel. It features an attractive sign for intra-sublattice scattering and a repulsive sign for inter-sublattice processes, in complete correspondence to the single layer\cite{Honerkamp2008}. The interlayer sign structure can be read off from the RG data as well. In detail, the leading part of effective interaction in this case (and ignoring frequency dependences) reads
$H_{\mathrm{AF}}=- \frac{1}{N} \sum_{o, o' }V_{oo'} \epsilon_o \epsilon_{o'}\vec{S}^{o}\cdot\vec{S}^{o'}$ with $V_{oo'}>0$ and $\vec{S}^{o}=\frac{1}{2} \sum_{\vec{k},s,s'}\vec{\sigma}_{ss^\prime} c^{\dagger}_{\vec{k},s,o} c_{\vec{k},s^\prime,o}$. The $\epsilon_o$s depend on the orbital, $\epsilon_o=+1$ for $o\in \{a_1,b_2\}$ and $\epsilon_o=-1$ for $o \in \{a_2,b_1\}$. The effective interaction has become infinitely-ranged due to the sharpness in momentum space.
A mean-field decoupling of $H_{\mathrm{AF}}$ results in an AF spin alignment in each layer where a net spin (e.g. 'up') moment is located on the A$_1$- and B$_2$-sublattices, and an opposite net spin ('down') moment on the B$_1$- and A$_2$-sublattices. In absence of spin-orbit interactions the spin quantization axis is not fixed. This phase is sometimes also referred to as a layer antiferromagnet (LAF) \cite{exp04}. A closer look at the fRG data shows that the intralayer components $V_{oo'}$ on the B sublattices grow faster toward the instability than the couplings on the A-sublattices, pointing to a larger spin moment on the B sublattice, in agreement with  Quantum Monte Carlo (QMC) simulations\cite{unserletter} for the same system with pure onsite interactions.
\\
{\em Charge density wave (CDW) instability.} Here we encounter diverging interactions in the density channel, again with zero momentum transfer, with opposite signs for the intra- and interorbital interactions. In detail, we observe the effective interaction
$
H_{\mathrm{CDW}}=-\frac{1}{N}\sum_{o, o'}V_{oo'} \epsilon_o  \epsilon_{o'} N^{o}N^{o'}        
$
with $V_{oo'}>0$ and $N^{o}=\sum_{\vec{k},s}c^{\dagger}_{\vec{k},s,o} c_{\vec{k},s,o}$.
Within a layer, this results in an infinitely-ranged attraction for sites on the same sublattice and repulsion for sites on different sublattices. The sign-structure between the layers favors an enhanced occupancy of the A$_1$ and B$_2$ sublattices and a reduced occupancy on the B$_1$ and A$_2$ sublattices or vice versa.  The electronic spectrum becomes  gapped by this ordering.
\\ 
{\em Quantum Spin Hall (QSH) instability.}
The QSH\cite{kanemele,qizhang} phase breaks spin-rotational symmetry, whereas time reversal symmetry remains conserved. In the fRG flow, spin interactions with zero wavevector transfer diverge, with an additional sign structure that alternates between $K$ and $K'$ points, and between the sublattices. The effective interaction reads
$
H_{\mathrm{QSH}}=- \frac{1}{N} \sum_{o, o'}V_{oo'} \epsilon_o \epsilon_{o'}\vec{S}^{o}_f\cdot\vec{S}^{o'}_f
$
with $V_{oo'}>0$ and $\vec{S}^{o}_f=\frac{1}{2}  \sum_{\vec{k},s,s'}f_{\vec{k}}\vec{\sigma}_{ss^\prime} c^{\dagger}_{\vec{k},s,o} c_{\vec{k},s^\prime,o}$ including a $f$-wave form factor $f_{\vec{k}}=\sin(k_x)-2 \sin(\frac{k_x}{2}) \cos(\frac{k_y \sqrt{3}}{2})$.  In a mean-field treatment of this effective interaction, a purely imaginary Kane-Mele\cite{kanemele} order parameter $\vec{\Delta}_{i,j}=\sum_{s,s'} \sigma_{ss'} \langle a^{\dagger}_{i,s} a_{j,s'} \rangle=\vec{\Delta}^*_{j,i}$ with next-nearest neighbors $i,j$ is induced. In the fRG, the chirality of the state comes out the same in the two layers for the same spin, i.e. there should be two edge modes with the same propagation direction per spin. Hence, the edge state would not be topologically protected\cite{qizhang}.
\\ 
{\em Three-sublattice CDW instability (CDW$_3$).}
Interestingly, we found another instability for smaller $U$ and $V_2/t\lesssim 1.0$, which so far seems not to be mentioned in the literature. Here a site-centered CDW tendency with a finite momentum transfer $\vec{Q}=\vec{K}-\vec{K}'=\vec{K}'$ grows during the fRG flow. The wavevector dependence of $V_\Lambda$ near $\Lambda_c$ for this CDW$_3$-instability is shown in Fig. \ref{fig:realspace-mcdw}. The sharp features belong to wavevector transfer $\pm \vec{Q}$. Only processes with initial $k_1$ and $k_2$ near different BCPs grow strongly, because only then the final states after scattering by $\pm \vec{Q}$  lie near the BCPs as well. This causes the interruption of the horizontal features in Fig.~\ref{fig:realspace-mcdw}. The corresponding effective interaction given by these leading terms near the divergence becomes 
\begin{eqnarray}
H_{\mathrm{CDW_3}}&=&- \frac{1}{N} \sum_{o,o'}V_{oo'} \epsilon_o \epsilon_{o'} \big(N^{o  }_{\vec{Q}} N^{o'}_{-\vec{Q}}+N^{o}_{-\vec{Q}} N^{o'}_{\vec{Q}}\big)
\label{hcdw3} 
\end{eqnarray}
with $N^{o}_{\vec{Q}}=\sum_{\vec{k},s}c^{\dagger}_{\vec{k}+\vec{Q},s,o} c_{\vec{k},s,o}$.
In this equation it is understood that the wavevectors $\vec{k}$ range in the vicinity of the $K$ and $K'$- points, as only there the interactions grow large. In a variational treatment, (\ref{hcdw3}) is minimized by complex expectation values $\langle N^{o}_{\vec{Q}} \rangle = \epsilon_o |\Delta_o| e^{i \alpha}Ê$. These break the translational symmetry by density modulations $\propto \cos ( \vec{Q} \cdot \vec{R} + \alpha)$. Based on the fRG data for $V_{oo'}$, the $|\Delta_o|$ should be of comparable magnitude. Each original sublattice $o$ is broken up into three sublattices (see Fig. \ref{fig:realspace-mcdw}). Changing $\alpha$ reorganizes the charges within the three new sublattices while keeping the average constant.  The discrete rotational symmetry is broken completely for general $\alpha$. Hence this state should be observable directly in scanning tunneling experiments. The fermionic spectrum is gapless and has 4 Dirac cones near the center of the new reduced BZ.
\\ \indent
Let us now discuss the relation to bilayer-graphene.  In Ref. \cite{wehling2011} the interaction parameters for single-layer graphene and graphite were estimated by ab-initio methods. We expect bilayer graphene to interpolate between these cases.
The area of these ab-initio interaction parameters  is shown as a dashed line in Fig. \ref{fig:Pd01}(d). In the fRG for this parameter range SDW and QSH instabilities compete. Which one is leading depends however on the precise values of the interaction and also the hopping parameters, so that we refrain from giving a definite prediction. We have also checked that a third-n.n. repulsion $V_3<t$ does not change the nature of the ground state for these parameters.  As both states have a non-zero single-particle-gap, they are compatible with the experimental spectrum of Ref. \cite{exp04}. A distinction might be made from testing for time-reversal symmetry breaking, which only occurs for the SDW state. As the QSH state corresponds to two copies of the Kane-Mele single-layer state, gapless edge state transport might be spoiled by impurities\cite{qizhang}. Further, for the $f$-wave order parameter of the QSH, impurities will be detrimental, and the experimental gap is only seen in ultraclean samples\cite{exp04}. 
\begin{figure}[t!]
\centering
\includegraphics[height=0.17\textwidth]{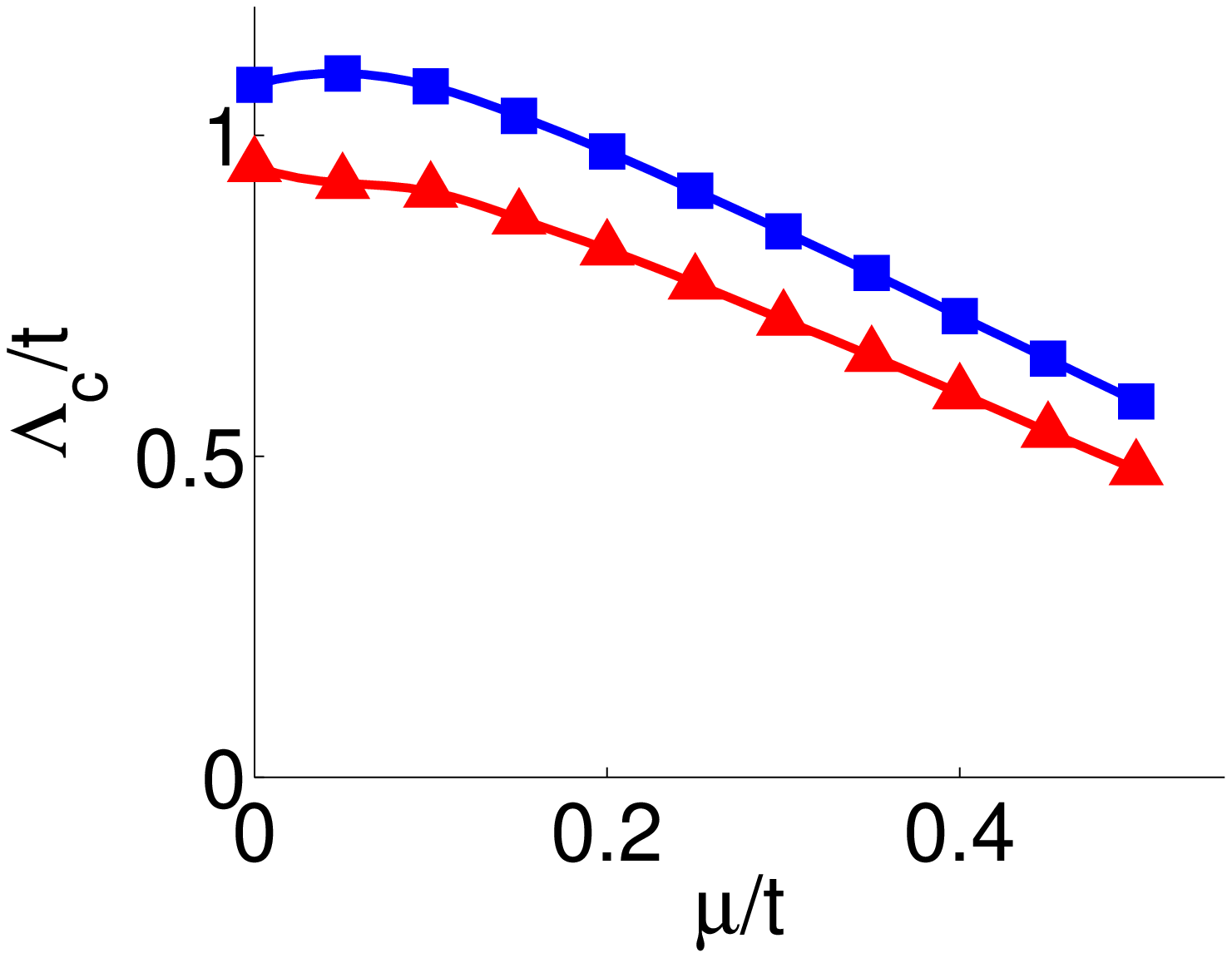}
\includegraphics[height=0.17\textwidth]{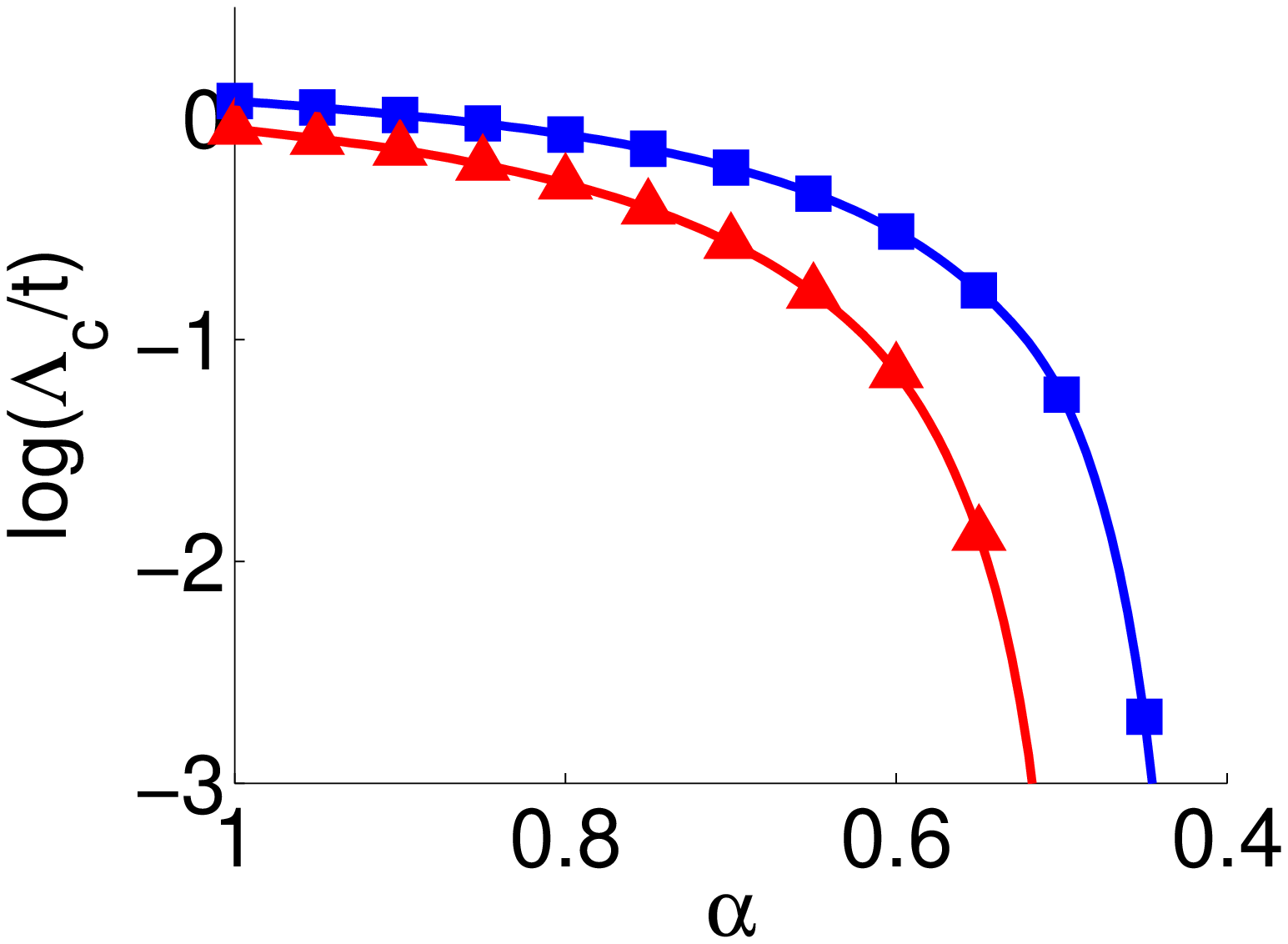}
\caption{Left panel: fRG critical scale $\Lambda_c$ versus chemical potential $\mu$ for graphene (squares) and  graphite (triangles) cRPA interaction parameters\cite{wehling2011}. Right panel: Critical scale $\Lambda_c$ as function of a rescaling parameter $\alpha$ for graphene (squares) and graphite (triangles) cRPA interaction parameters. }
\label{fig:muplot}
\end{figure}
\\ \indent
We also investigated the impact of non-vanishing chemical potentials ($\mu \neq 0$) to mimic the effect of impurities or small dopings on the groundstate, with focus on the cRPA interaction parameters and $t_\perp=0.1 t$. In the range between $\mu\in (0,0.5 t]$ the critical scale $\Lambda_c$ only changes mildly and the groundstate remains unchanged. Note that the instability scales deduced for these realistic parameters are huge (up to $\sim t$), due to the perfect nesting between the two bands forming the quadratic BCPs. We cannot guarantee that our method still works quantitatively in this regime. Most likely the inclusion of self-energy corrections and frequency dependence of the interactions will reduce the fRG scales. However, we show this data on purpose in order to expose clearly that the experimental energy gaps\cite{exp04} $\sim 2\, \mathrm{meV} \sim 10^{-3}t$ are several orders of magnitude smaller than the gap scales one gets theoretically in this simple modeling, using a method that should be expected to be more realistic than mean-field theory or simpler perturbative arguments. 
The single-layer system for pure onsite interactions offers a possibility to compare the energy scales in the fRG  with published non-perturbative QMC data.  At $U\approx 4.3t$ where the AF order sets in in QMC\cite{meng} above a narrow spin-liquid regime, the single-particle gap is $\sim 0.15t$. The temperature-flow fRG\cite{Honerkamp2008} with upper bandwidth $\sim \pm t$ gave $\Lambda_c \sim 0.16t$, with a critical $U_c\sim 3.8t$ for AF order. Our present code with energy-cutoff and integration over the full bandwidth gives a higher $\Lambda_c=0.85t$ at $U=4.3t$, which is already far above the corresponding $U_c=2.8t$ within this scheme. Hence the overestimate may reach factors $\sim 5-6$, but  in a limited range of parameters near the opening of the gap at $U_c$ in QMC. In the bilayer system, $U_c$ is expected to be zero and high scales occur for a wide range of parameters.  We expect that other methods will confirm this fRG result.
Of course, in the experimental system, unintentional doping and potential variations will lead to a reduction of the energy scales. But as one can see from the doping dependence just described, a single perturbation has to be rather strong to be effective, and one may actually need a combination of factors like doping, disorder, trigonal warping, additional reductions of the interaction parameters etc., to reduce the scales down to values compatible with the experiment. In order to estimate the dependence of the ground state on the critical scale, we introduced a global rescaling parameter $\alpha$ for the interaction terms, i.e. $U\rightarrow \alpha U$, $V_1\rightarrow \alpha V_1$, $V_2\rightarrow \alpha V_2$. We found that $\alpha$ does not change the nature of the ground state, i.e. for the graphene parameters we observed a QSH instability for all $\alpha$ and for the graphite parameters the SDW instability. $\Lambda_c$ decreases by two orders of magnitude as $\alpha\lesssim 0.5$, as shown in the right plot of Fig. \ref{fig:muplot}. The marked reduction is caused by the strongly energy-dependent density of states.  
\\ \indent
In summary, we have presented a fRG study of interaction driven instabilities in the honeycomb bilayer model. Besides a novel gapless CDW state, we found that using ab-initio estimates for the band-structure and non-local interaction parameters for bilayer graphene leads to a narrow competition of quantum-spin-Hall and AF-SDW instabilites, where details might decide what the actual groundstate is. Another important information is the typically high energy scale for the breakdown of the gapless state in the model used by us and many other theoretical studies. At present, these high scales do not seem to be reflected in the experiments, and more research is needed to understand this discrepancy.
\\ \indent
We acknowledge discussions with S. Bl\"ugel, T. C. Lang, M. J. Schmidt, T. Wehling, and S. Wessel and support by DFG research units FOR 723, 912 and 1162.

\end{document}